\documentclass[12pt]{article}

\usepackage{amsfonts,amssymb,amsmath,enumerate}
\usepackage{color}
\usepackage{theorem,cite}
\usepackage{indentfirst}
\usepackage{textcomp}

\theorembodyfont{\rmfamily}

\newcommand{\bea}{\begin{eqnarray}}
\newcommand{\eea}{\end{eqnarray}}
\newcommand{\shit}{\begin{align}}
\newcommand{\mdr}{\end{align}}
\newcommand{\beano}{\begin{eqnarray*}}
\newcommand{\eeano}{\end{eqnarray*}}
\newcommand{\beq}{\begin{equation}}
\newcommand{\eeq}{\end{equation}}

\newcommand{\mb}[1]{\hspace{2.1ex}\mbox{#1}\hspace{2.1ex}}
\numberwithin{equation}{section}

\newcommand{\eps}{\epsilon}

\def\ft{{\mathfrak t}}

\def\cS{{\cal S}}    \def\cT{{\cal T}}

\newcommand{\CC}{{\mathbb C}}

\newcommand{\II}{{\mathbb I}}

\newcommand{\wt}[1]{\widetilde{#1}}

\newcommand{\id}{{\II}}

\begin{document}
\pagestyle{empty}
\begin{flushright}
LAPTH-055/15
\end{flushright}

\vspace{20pt}

\begin{center}
\begin{LARGE}
{\bf 3-state Hamiltonians associated}\\[1.2ex]
{\bf to solvable $33$-vertex models}
\end{LARGE}

\vspace{50pt}

\begin{large}
{N.~Cramp\'e${}^{a}$, L.~Frappat${}^b$, E.~Ragoucy${}^b$,  M. Vanicat${}^b$ \footnote[1]{ncrampe@um2.fr, luc.frappat@lapth.cnrs.fr, eric.ragoucy@lapth.cnrs.fr, matthieu.vanicat@lapth.cnrs.fr}}
\end{large}

\vspace{15mm}

${}^a$ {\it Laboratoire Charles Coulomb (L2C), UMR 5221 CNRS-Universit\'e de Montpellier,\\
Montpellier, F-France.}

\vspace{5mm}

${}^b$ {\it Laboratoire de Physique Th\'eorique LAPTH, CNRS and Universit\'e de Savoie,\\
BP 110, 74941 Annecy-le-Vieux Cedex, France}

\end{center}

\vspace{4mm}

\begin{abstract}
Using the nested coordinate Bethe ansatz, we study 3-state Hamiltonians with 33 non-vanishing entries, 
or $33$-vertex models, where only one global charge with degenerate eigenvalues exists and
each site possesses three internal degrees 
of freedom. In the context of Markovian processes, 
they correspond to diffusing particles with two possible internal states which may be exchanged during the diffusion 
(transmutation).
The first step of the nested coordinate Bethe ansatz is performed providing the eigenvalues in terms of rapidities. 
We give the constraints ensuring the consistency of the computations. These rapidities also satisfy Bethe equations
involving $4\times 4$ $R$-matrices, solutions of the Yang--Baxter equation which implies new constraints on the models. 
We solve them allowing us to list all 
the solvable $33$-vertex models.
\end{abstract}


\newpage
\pagestyle{plain}
\setcounter{page}{1}

\section{Introduction}

The analytical computation of the eigenvalues of a quantum Hamiltonian is a rather difficult question, that is even usually impossible to achieve.
However, in \cite{Bethe}, H. Bethe succeeded in computing exactly the spectrum of the one-dimensional Heisenberg quantum spin chain \cite{Heis}.
Since his work, the classification of solvable one-dimensional systems has been the heart of a lot of researches.
General methods are now known but involve huge computations 
which, in general, do not permit to provide a classification of solvable models.

In this letter, we are interested in finding  Hamiltonians $H$ such that the following eigenvalue problem
\begin{equation}\label{eq:EP}
 H\Psi =E \Psi
\end{equation}
can be solved exactly. The Hamiltonians under consideration correspond to
 one-dimensional spin chains with nearest neighbor interactions,  and are written as follows
\begin{equation}
 H=\sum_{\ell=1}^L h_{\ell,\ell+1}\;
\end{equation}
where we assume periodic boundary conditions (by convention $L+1\equiv 1$) and the indices $\ell,\ell+1$ indicate in which spins the local Hamiltonian $h$ acts. This type of problem also appears in the context of two-dimensional equilibrium statistical models or one-dimensional out-of-equilibrium models. The archetypes of such models are respectively the 6-vertex model \cite{baxter} 
or the asymmetric exclusion process \cite{ASEP1,ASEP2,ASEP3}.
Here, we focus on the case where $h$ is a $9\times 9$ matrix which means that each site may take three different values.

The cases when $H$ commutes with some charges of the following type
\begin{equation}\label{eq:charge}
 Q=\sum_{\ell=1}^L  \id_3^{\otimes \ell-1}\otimes q \otimes \id_3^{\otimes L-\ell}
\end{equation}
where $\id_p$ is the $p\times p$ identity matrix, are of particular interest. 
In the context of out-of-equilibrium models, the number of conserved charges 
corresponds to the number of type of particles (one of the value is for the empty site).

When $H$ commutes with two different charges, the number of non vanishing elements of $h$ is reduced to $15$ (in the context of statistical mechanics, 
it is called the 15-vertex model). In out-of-equilibrium statistical physics, it corresponds to models where there are two classes of conserved particles (such as two species ASEP).

When $H$ commutes with only one charge, one usually takes $q$ with three different eigenvalues: $h$ has then $19$ non vanishing entries. Such solvable models have 
been classified and studied previously in \cite{ITA, MP, Mart, CFR, FFR}.  They correspond to out-of-equilibrium models with only one species of particle, but where two particles can occupy the same site.

We concentrate here on Hamiltonians $H$ commuting with one charge that has one eigenvalue degenerated twice: 
$h$ has then $33$ non vanishing entries. 
Usually, the denomination $33$-vertex model is dedicated to integrable models whose $R$-matrix has 
$33$ non-vanishing entries. However, it appears that for the models solvable by CBA, when an $R$-matrix can be constructed, 
its non-vanishing entries coincide with the ones of $h$, see construction in \cite{FFR,nous}. 
Although this property has not been proven in full generality, but only checked case by case, we will  call our Hamiltonians, $33$-vertex Hamiltonians.
These models may be interpreted, in the context of out-of-equilibrium model, as diffusing particles possessing two internal degrees of freedom.  
Let us emphasize that such models  have been also introduced 
to study mRNA translation in \cite{bio}. Therefore, we hope that the solvable models introduced here may be helpful in this context
or to describe other phenomena.

Let us remark that the three cases described above exhaust all the non trivial cases when $H$ possesses conserved charge(s). 
To fix the notations, we introduce the canonical basis 
\begin{equation}\label{eq:bais}
 |1\rangle=\begin{pmatrix}1\\0\\0\\ \end{pmatrix}\quad,\qquad |2\rangle=\begin{pmatrix} 0\\1\\0\\ \end{pmatrix}\quad\text{and}\qquad 
|3\rangle=\begin{pmatrix} 0\\0\\1\\ \end{pmatrix}\;.
\end{equation}
The vector $|1\rangle$ will correspond to the empty site whereas $|2\rangle$ and $|3\rangle$ correspond to a 
particle in different internal states. In this context, the most general Hamiltonians which preserve the number of particles are the ones which commute with the 
charge \eqref{eq:charge} with
\begin{equation} \label{def:q}
q=\begin{pmatrix} 0 &0 &0\\ 0 &1 &0\\ 0 &0 &1\\ \end{pmatrix}\;.
\end{equation}
Therefore the local Hamiltonian $h$ takes the following form
\begin{equation}
h = 
\begin{pmatrix}
m_{11} & 0 & 0 & 0 & 0 & 0 & 0 & 0 & 0 \\
0 & m_{22} & m_{23} & m_{24} & 0 & 0 & m_{27} & 0 & 0 \\
0 & m_{32} & m_{33} & m_{34} & 0 & 0 & m_{37} & 0 & 0 \\
0 & m_{42} & m_{43} & m_{44} & 0 & 0 & m_{47} & 0 & 0 \\
0 & 0 & 0 & 0 & m_{55} & m_{56} & 0 & m_{58} & m_{59} \\
0 & 0 & 0 & 0 & m_{65} & m_{66} & 0 & m_{68} & m_{69} \\
0 & m_{72} & m_{73} & m_{74} & 0 & 0 & m_{77} & 0 & 0 \\
0 & 0 & 0 & 0 & m_{85} & m_{86} & 0 & m_{88} & m_{89} \\
0 & 0 & 0 & 0 & m_{95} & m_{96} & 0 & m_{98} & m_{99} \\
\end{pmatrix}.
\label{eq:ham12}
\end{equation}
Note that the $15$-vertex model is a sub-case of the problem studied here in opposition to the $19$-vertex model that exhibits different non vanishing entries.

The goal of this paper consists in classifying all such models which are solvable by Coordinate Bethe Ansatz (CBA).
 It is our main result, exposed in section \ref{sec:main}.
The proof is detailed in the following sections. In section \ref{sec:CBA} we perform the first step of the CBA (i.e. the nesting), which 
leads to a reduced problem dealing with $4\times4$ $R$-matrices possessing a specific form. In section \ref{sec:ybe} we classify these $R$-matrices, 
solutions of a braided Yang--Baxter equation with spectral parameters.

\section{Main result \label{sec:main}}
The calculations to get constraints on the entries of the Hamiltonian are detailed in sections \ref{sec:CBA} and \ref{sec:ybe}.
In this section, we just present the final result, after gathering the different constraints.  We obtain the following classification of 33-vertex Hamiltonians.
The Hamiltonian \eqref{eq:ham12}
is solvable by CBA if and only if its entries obey the following constraints:
\begin{enumerate}
\item[$(i)$] We must have:
\begin{equation}\begin{aligned}
& m_{23}+m_{47}=0\quad&;\qquad m_{32}+m_{74}=0\quad&;\qquad m_{27}=m_{34}=m_{43}=m_{72}=0\\
& m_{24}=m_{37}\quad&;\quad \qquad m_{42}=m_{73}\qquad&;\qquad m_{22}+m_{44}=m_{33}+m_{77},
\end{aligned}
\label{1st.constraint}
\end{equation}
\item[$(ii)$] The matrix
\beq\label{eq:Tmat}
T= \begin{pmatrix}t_{55} & t_{56} & t_{58} & t_{59} \\
 t_{65} & t_{66} & t_{68} & t_{69} \\
 t_{85} & t_{86} & t_{88} & t_{89} \\
 t_{95} & t_{96} & t_{98} & t_{99} \end{pmatrix}
\eeq
where
\beq\label{eq:m.tilde}
\begin{aligned} 
t_{55} = m_{55} +m_{11} - m_{22} - m_{44} \ ;\quad
t_{58} = m_{58} +m_{23} \ ;\quad t_{85} = m_{85} + m_{32}\ ;\quad t_{59} = m_{59}\\
t_{66} = m_{66} +m_{11} - m_{33} -  m_{44} \ ;\quad
t_{69} = m_{69} +m_{23} \ ;\quad t_{96} = m_{96} + m_{32}\ ;\quad t_{68} = m_{68}\\
t_{88} = m_{88} +m_{11} - m_{22} - m_{77}\ ;\quad
t_{56} = m_{56} -m_{23} \ ;\quad t_{65} = m_{65} - m_{32}\ ;\quad t_{86} = m_{86}\\
t_{99} = m_{99} +m_{11} - m_{33} - m_{77}\ ;\quad
t_{89} = m_{89} -m_{23} \ ;\quad t_{98} = m_{98} - m_{32} \ ;\quad t_{95} = m_{95}
\end{aligned}
\eeq
must be a representation of one of these three algebras:
\begin{enumerate}
\item \textbf{Hecke algebras}
\bea\label{eq:toto}
T_{12}T_{23}T_{12}-m_{24}m_{42}\,T_{12}=T_{23}T_{12}T_{23}-m_{24}m_{42}\,T_{23}\mb{and}
T^2=\mu\,T
\eea
\item \textbf{$\cT_n$ algebras}
\beq \label{eq:titi}
T_{12}T_{23}T_{12}+T_{12}\, (T_{23})^2=T_{23}T_{12}T_{23}+(T_{12})^2\, T_{23}
\eeq
\item \textbf{$\cS_n$ algebras}
\beq \label{eq:tata}
T_{12}T_{23}T_{12}+ (T_{23})^2\, T_{12}=T_{23}T_{12}T_{23}+T_{23}\,(T_{12})^2
\eeq
\end{enumerate}
Fortunately, the $4\times4$ solutions to relations \eqref{eq:toto}, \eqref{eq:titi} or \eqref{eq:tata} have been classified: all the possible expressions of $T$ can be then recovered from these classifications. 

For Hecke algebras, setting $\widetilde T=(T - \rho)/\tau $ with $\tau^2=\mu\rho$ and $\rho^2-\mu\rho+m_{24}m_{42}=0$, $\widetilde T$ obeys 
in particular the Braided Yang--Baxter equation, whose $4\times4$ solutions have been classified in \cite{Hiet}. The second relation in \eqref{eq:toto} provides additional constraints on $\wt T$. 
 Taking the same notation as \cite{Hiet}, it implies that it
must be chosen among the 7 matrices $R_{H2,1}$, $R_{H2,2}$, $R_{H1,1}$, $R_{H1,2}$, $R_{H1,3}$, $R_{H1,4}$, $R_{H0,3}$ subjected to the constraints given in  table \eqref{hieta}.

The $\cS_n$ and $\cT_n$ algebras have been introduced in \cite{nous} to define a new type of Baxterisation. The classification of the $4\times4$ matrices corresponding to $\cS_n$ or $\cT_n$ can be found there.

\end{enumerate}

We present now the proofs of this result. Constraint $(i)$ is found in section \ref{sec:CBA} and  constraint $(ii)$ is obtained in section \ref{sec:ybe}.

\section{Coordinate Bethe ansatz\label{sec:CBA}}

As mentioned above, we now focus on Hamiltonians \eqref{eq:ham12} that commute with $Q$, defined by \eqref{eq:charge} and \eqref{def:q}. 
The basis \eqref{eq:bais} allows us to introduce the following elementary states
\begin{equation}\label{eq:elemstate}
 \vert x_1,n_1;x_2,n_2;\dots;x_M,n_M \rangle= |1\rangle^{\otimes x_1-1} \otimes |n_1\rangle\otimes |1\rangle^{\otimes x_2-x_1-1}\otimes |n_2\rangle
 \dots \otimes |n_M\rangle \otimes |1\rangle^{\otimes L-x_M}
\end{equation}
where $n_i=2,3$ and $x_i=1,2,\dots,L$. 
In words, the state $\vert x_1,n_1;x_2,n_2;\dots;x_M,n_M \rangle$ stands for the configuration where the particles are in positions $x_1, x_2,\dots,x_M$ 
with internal states $n_1,n_2,\dots n_M$ respectively\footnote{We remind that $|1\rangle$ stands for an empty site.}. 

Notice that $\vert x_1,n_1;x_2,n_2;\dots;x_M,n_M \rangle$ is an eigenvector of $Q$ with eigenvalue $M$ whatever the values of $x_i$'s and $n_i$'s are. In fact, the set of states $\vert x_1,n_1;x_2,n_2;\dots;x_M,n_M \rangle$ spans the vector space with $M$ particles.
Therefore, an Hamiltonian eigenstate $\Psi_M$  in a given sector with $M$ particles can be written as a linear combination 
of the elementary states \eqref{eq:elemstate} with coefficients $a(x_1,\dots,x_M)$, which are complex-valued functions to be determined:
\begin{equation}\label{eq:psiM}
\Psi_M = \sum_{1 \leq x_1 < \dots < x_M \le L}\ \sum_{n_1,n_2,\dots,n_M=2,3}\  a_{n_1,n_2,\dots,n_M}(x_1,x_2,\dots,x_M)  \vert x_1,n_1;x_2,n_2;\dots;x_M,n_M \rangle.
\end{equation}
The coordinate Bethe ansatz \cite{Bethe} consists in assuming a plane wave decomposition for the functions $a_{n_1,n_2,\dots,n_M}(x_1,x_2,\dots,x_M)$:
\begin{equation}
a_{n_1,n_2,\dots,n_M}(x_1,x_2,\dots,x_M) = \sum_{\sigma \in {\mathfrak S}_M} A_\sigma^{(n_1,n_2,\dots,n_M)} \prod_{n=1}^M z_{\sigma(n)}^{x_n} ,
\label{eq:planewave}
\end{equation}
where ${\mathfrak S}_M$ is the permutation group of $M$ elements.
The unknowns $A_\sigma^{(n_1,n_2,\dots,n_M)}$ are functions on the symmetric group algebra depending 
on the parameters $z_1, z_2, \dots, z_M$ called rapidities 
and which are solutions of the Bethe equations determined below.
To simplify the following computations, we encompass the $2^M$ different unknowns $A_\sigma^{(n_1,n_2,\dots,n_M)}$ for a given $\sigma$ in the following
vector
\begin{equation}
 \boldsymbol{A_\sigma}=\sum_{n_1,n_2,\dots,n_M=2,3} A_\sigma^{(n_1,n_2,\dots,n_M)}  
 |\overline{n}_1\rangle\otimes  |\overline{n}_2\rangle \otimes \dots \otimes  |\overline{n}_M\rangle
\end{equation}
where
\begin{equation}\label{eq:basis}
\qquad |\overline{2}\rangle=\begin{pmatrix} 1\\0\\ \end{pmatrix}\quad\text{and}\qquad 
|\overline{3}\rangle=\begin{pmatrix} 0\\1\\ \end{pmatrix}\;.
\end{equation}

As usual, we project the eigenvalue problem \eqref{eq:EP} on the different elementary states, the eigenvector having the form \eqref{eq:psiM} 
with \eqref{eq:planewave}. We do not detail the calculations, since they are similar to the nested coordinate Bethe ansatz developed in \cite{Yang,Suth}
based on the ideas of \cite{Bethe}. The computations are divided into two main steps:
\begin{itemize}
 \item we reduce the original eigenvalue problem with $L$ sites allowing three different states, to an eigenvalue problem for a system with a smaller number of sites, that allow only two different states
 (this system is called the reduced problem);
 \item we determine (and classify) when the  reduced problem is integrable. 
\end{itemize}
In the following, we  sketch these two steps and give the  main results.

\paragraph{First step.}
Performing the CBA on the position of the particles (and not looking at their internal states), we get a first set of constraints 
on the parameters of the local Hamiltonian. This corresponds to the constraints \eqref{1st.constraint} given in section \ref{sec:main}.
This first step allows us to determine the energy of the state $\Psi_M$:
\begin{equation}
E_M = L \, m_{11} + \sum_{n=1}^M \epsilon(z_n) \qquad \text{with} \qquad 
\epsilon(z) = m_{22}+m_{44}-2m_{11} + \frac{m_{24}}{z} + m_{42}\, z
\label{eq:energie}
\end{equation}
provided the coefficients $\boldsymbol{A_\sigma}$ 
are related by
\begin{align}
& \boldsymbol{A_{\sigma \ft_j}}= \check S_{j,j+1}(z_{\sigma(j)},z_{\sigma(j+1)})\,\boldsymbol{A_{\sigma}} \label{eq:AAS}\\
& \check S(z_1,z_2) = -\frac{z_2}{z_1}\,\Lambda(z_1,z_2)\,\Lambda(z_2,z_1)^{-1}\,,\label{eq:Smatrix}
\end{align}
where $\ft_j \in {\mathfrak S}_M$ denotes the transposition $(j,j+1)$ and
\begin{align}\label{def:Lambda}
&\Lambda(z_1,z_2) = T-\left(m_{42}\, z_1+\frac{m_{24}}{z_2}\right)\,\II_4 .
\end{align}
$T$  is the $4\times4$ constant matrix \eqref{eq:Tmat} whose entries depend on the entries of $h$, as stated in \eqref{eq:m.tilde}.  

Let us describe more precisely the meaning of the indices of $\check S_{j,j+1}$ in \eqref{eq:AAS}: they indicate in which spaces the $4\times4$ 
matrix $\check S$ acts non trivially in the tensor product $(\CC^2)^{\otimes M}$ spanned by 
$\{|\overline{n}_1\rangle\otimes  |\overline{n}_2\rangle \otimes \dots \otimes  |\overline{n}_M\rangle\ |\ n_i=2,3\}$.
 Explicitly, we get
 \begin{equation}
  \check S_{j,j+1}(z_j,z_{j+1})=\id_2^{\otimes j-1}\otimes \check S(z_j,z_{j+1})\otimes \id_2^{\otimes L-j-1}\;.
 \end{equation}

Due to the defining relations of the permutation group ${\mathfrak S}_M$ 
$$\ft_j^2=id,\qquad [\ft_j,\ft_k]=0,\ |k-j|>1,\qquad \ft_j\ft_{j+1}\ft_j=\ft_{j+1}\ft_j\ft_{j+1},$$
relations \eqref{eq:AAS} gives constraint on $\check S$:
\begin{eqnarray}
 &&\check S_{j,j+1}(z_j,z_{j+1})\check S_{j,j+1}(z_{j+1},z_j)=1\quad,\qquad \big[ \check S_{j,j+1}(z_j,z_{j+1}), \check S_{k,k+1}(z_k,z_{k+1}) \big] = 0 \qquad \\
 &&\check S_{12}(z_1,z_2) \check S_{23}(z_1,z_3) \check S_{12}(z_2,z_3)=\check S_{23}(z_2,z_3) \check S_{12}(z_1,z_3)\check S_{23}(z_1,z_2).\label{eq:YBE}
\end{eqnarray}
The first two relations are trivially satisfied by \eqref{eq:Smatrix}. The third one \eqref{eq:YBE}, called braided Yang--Baxter equation, holds only
if supplementary constraints on the entries of $T$ are satisfied. We postpone the study of these constraints in section \ref{sec:ybe} and
suppose from now on that they are indeed satisfied.
  
Because of the periodicity of the model, the rapidities $z_j$ are quantified and must obey the first set of Bethe equations 
\begin{equation}
z_{j}^{L}\, \boldsymbol{A_{id}}= 
 S_{j+1,j}(z_{j+1},z_j)\cdots S_{M,j}(z_M,z_j)\, S_{1,j}(z_{1},z_j)\cdots S_{j-1,j}(z_{j-1},z_j)\,\boldsymbol{A_{id}}\,, 
\quad j=1,...,M.
\label{eq:BAE}
\end{equation}
where
\begin{equation}\label{S-Scheck}
 S(x,y)= P\check S(x,y) \quad\text{and}\qquad 
 P= \begin{pmatrix}1 &0 & 0& 0 \\
 0 & 0 & 1 & 0 \\ 0 & 1 & 0 & 0 \\ 0 & 0 & 0 & 1 \end{pmatrix}.
\end{equation}
This set of eigenvalue problems is the reduced problem. The matrix $S$ is a $4\times 4$ matrix.

\paragraph{Second step: nesting.}
The matrix $ S(x,y)$ is obviously regular, $S(x,x)=-P$, 
so that the set of eigenvalue problems \eqref{eq:BAE} can be recasted using a transfer matrix
\begin{equation}\label{eq:tran}
t(z;\{z_1,...z_M\})=tr_0\Big(S_{10}(z_1,z)\cdots  S_{M0}(z_M,z)\Big)\;.
 \end{equation}
Because of the Yang--Baxter equation \eqref{eq:YBE}, the transfer matrix commute for different values of $z$:
\begin{equation}
 \big[ t(x;\{z_1,...z_M\}),t(z;\{z_1,...z_M\}) \big] = 0\;.
\end{equation}

Therefore, the Bethe equations \eqref{eq:BAE} reduced to  
\begin{equation}
z_{j}^{L}\, \boldsymbol{A_{id}}=-t(z_j;\{z_1,...z_M\})\boldsymbol{A_{id}}
\end{equation}
are compatible since we can diagonalize simultaneously all the $t(z_j;\{z_1,...z_M\})$.
Let us remark that the transfer matrix \eqref{eq:tran} may be used to define a new integrable system.

To finish the computations, one should solve the reduced problem using, for example, the Bethe ansatz once again.
However, in the cases treated here, the reduced problem has no conserved charge and the resolution becomes much harder.    
How to apply the Bethe ansatz in these cases is still an open question. However, let us mention that new methods appeared recently in order to solve similar problems where there is no conserved 
charge due to the boundary conditions (generalization of the CBA \cite{Dam}, Onsager approach \cite{pascal}, separation of variables \cite{nic},
inhomogeneous Bethe equation \cite{CYSW}, modified algebraic Bethe ansatz \cite{sam}). 
A generalization of these methods may be possible to deal with the eigenvalue problem \eqref{eq:BAE}.  
In the case of Markovian processes, the matrix ansatz developed in \cite{DEHP} with its link with integrability \cite{SW,CRV} may be also helpful
for the resolution of this eigenvalue problem.

\section{Braided Yang--Baxter equation\label{sec:ybe}}

This section is devoted to the classification of the matrices $T$ such that the braided Yang--Baxter equation \eqref{eq:YBE} holds.
We split the problem into three subcases: $i$) $m_{24}m_{42}\neq 0$, then $ii$) $m_{42}=0, m_{24}\neq 0$ and $iii$) $m_{24}=0, m_{42}\neq 0$. The case $m_{24}=m_{42}= 0$ is
excluded since it corresponds to an energy \eqref{eq:energie} which does not depend on the rapidities (there is no diffusion of particles).

\subsection{Case $\boldsymbol{m_{24}\,m_{42}\neq 0}$\label{sec:Hecke}} 
By taking different expansions w.r.t. $z_1$, $z_2$ and $z_3$ 
 in \eqref{eq:YBE} and after algebraic manipulations,
we find that the braided Yang--Baxter equation \eqref{eq:YBE} holds if and only if $T$ satisfies:
\begin{equation}
\label{eq:heckem1}
\begin{split}
 & T_{12}T_{23}T_{12}-m_{24}m_{42}T_{12}=T_{23}T_{12}T_{23}-m_{24}m_{42}T_{23}\,, \\
 & (T_{12})^2\ T_{23}=T_{12}\ (T_{23})^2\,, \\
 & (T_{23})^2\ T_{12}=T_{23}\ (T_{12})^2\;.
\end{split}
\end{equation}
Note that these relations come directly from the form \eqref{eq:Smatrix} of the $S$-matrix, and do not depend on the size of $T$. 
In general, classifying the solutions of equations \eqref{eq:heckem1} is a difficult task. However, in the case of $4\times4$ matrices treated here, 
it is  possible to use formal mathematical software to deal with them. Firstly, we prove that all the solutions of \eqref{eq:heckem1}
verify
\begin{equation}\label{eq:mu}
 T^2= \mu T\;.
\end{equation}
Secondly, we perform the following transformation
\begin{equation}\label{eq:T-Ttilde}
 T=\tau \widetilde T + \rho
\end{equation}
with $\rho$ solution of $\rho^2-\mu\rho+m_{24}m_{42}=0$ and $\tau=\pm\sqrt{\rho^2+m_{24}m_{42}}$.
The matrix $\widetilde T$ satisfies the relations of the Hecke algebra
\begin{eqnarray}
&& \widetilde T_{12}\widetilde T_{23}\widetilde T_{12}=\widetilde T_{23}\widetilde T_{12}\widetilde T_{23}\quad,\\
&& \widetilde T-\widetilde T^{-1}=\frac{\mu-2\rho}{\tau}\equiv \wt\mu\;.\label{eq:hecke}
\end{eqnarray}
Thirdly, we pick up, in the classification of the constant braided Yang--Baxter equation done in \cite{Hiet}, the $4\times 4$ matrices satisfying also the relation \eqref{eq:hecke}.

The classification of \cite{Hiet} provides 23 solutions, up to transformations
\beq\label{transfo:hecke}
\wt T \mapsto \lambda\, g \otimes g\ \wt T\ g^{-1}\otimes g^{-1} \mb{;}\wt T \mapsto \wt T^{t_1t_2} \mb{and} \wt T \mapsto P\,\wt T\,P,
\eeq
where $\lambda$ is a complex non-zero parameter, $g$ is any invertible $2\times2$ matrix and $P$ is the permutation matrix, see eq. \eqref{S-Scheck}.
Among these solutions, only 9 are also solution to \eqref{eq:hecke} and this supplementary relation imposes (in general) constraints on their parameters.
Using the notations of \cite{Hiet}, these solutions are (up to the transformations  \eqref{transfo:hecke}):
\begin{equation}\label{hieta}
\begin{tabular}{| c | c | c |}
\hline
 \rule{0mm}{2.5ex} Matrix & Constraints on the parameters & Value of $\wt{\mu}$ \\[.7ex] \hline 
 \rule{0mm}{2.4ex} $R_{H3,1}$ & $pq=1$, $\quad k^2=1$, $\quad s^2=1$ & $0$ \\[.4ex]  \hline 
 \rule{0mm}{2.4ex} $R_{H2,1}$ & $k^2pq=1$ & $k^2-pq$ \\[.4ex] \hline 
 \rule{0mm}{2.4ex} $R_{H2,2}$ & $k^2pq=1$ & $k^2-pq$ \\[.4ex] \hline 
 \rule{0mm}{2.4ex} $R_{H2,3}$ & $k^2=1$, $\quad p+q=0$, $\quad 2ks+p^2+q^2=0$ & $0$ \\[.4ex] \hline
 \rule{0mm}{2.4ex} $R_{H1,1}$ & $4p^2q^2=1$ & $2(p^2-q^2)$ \\[.4ex] \hline
 \rule{0mm}{2.4ex} $R_{H1,2}$ & $pq=1$ & $p-q$ \\[.4ex] \hline 
 \rule{0mm}{2.4ex} $R_{H1,3}$ & $k^4=1$ & $0$ \\[.4ex] \hline 
 \rule{0mm}{2.4ex} $R_{H1,4}$ & $pq=1$, $\quad k^2=1$ & $0$ \\[.4ex] \hline 
 \rule{0mm}{2.4ex} $R_{H0,3}$ & no constraint & $0$ \\[.4ex] 
 \hline 
\end{tabular}
\end{equation}

 After imposing these constraints, the previous matrices are not independent anymore: they can be all 
obtained  from the 7 matrices $R_{H2,1}$, $R_{H2,2}$, $R_{H1,1}$, $R_{H1,2}$, $R_{H1,3}$, $R_{H1,4}$, $R_{H0,3}$ 
(subjected to the constraints given in  \eqref{hieta}). 
The value of $\mu$ in \eqref{eq:mu} is reconstructed from these data. Indeed, simple algebraic manipulations show that 
\beq\label{eq:mutau}
\mu^2 = \frac{m_{24}\,m_{42}} {x^2(1-x^2)}\,,\quad  \rho=\mu\,x^2\,,\quad \tau =x\,\mu 
 \mb{where $x$ is a solution of}  2x^2+\wt\mu\, x-1 = 0.
\eeq

Finally, we remark that for any previous solutions of the Hecke algebra, the matrix $\check S$ verifies the Yang--Baxter equation \eqref{eq:YBE}. 
Performing the change of variable
\begin{equation}
z_j=\frac{\mu(1-x_j)+\delta(1+x_j)}{2m_{42}(1-x_j)}, \qquad \mbox{with} \quad \delta=\sqrt{\mu^2-4m_{24}m_{42}},
\end{equation}
one can verify that $\check S(z_1,z_2)$ depends only on the ratio $\dfrac{x_1}{x_2}$ (up to a normalisation factor).
In fact, in terms of the variables $x_j$,  equation \eqref{eq:Smatrix} is equivalent to the usual Baxterisation of the Hecke algebra introduced in \cite{Jo} (see also \cite{FFR}).   

\subsection{Case $\boldsymbol{m_{24}=0}$ and  $\boldsymbol{m_{42}\neq 0}$.}
Now, the different expansions w.r.t. $z_1$, $z_2$ and $z_3$ 
 in \eqref{eq:YBE}, lead to the sole relation:
\begin{eqnarray}\label{eq:m24=0}
T_{12}T_{23}T_{12}+T_{12}\ (T_{23})^2=T_{23}T_{12}T_{23}+(T_{12})^2\ T_{23}
\end{eqnarray}

One recognizes in \eqref{eq:m24=0} the algebra $\cT_3$ introduced in \cite{nous} to define a new Baxterisation for $R$-matrices depending separately on two spectral parameters. The $4\times 4$ solutions have been classified there. There are 7 classes of solutions, up to the group of transformations generated by
\beq\label{eq:transfoTn}
T \mapsto \lambda\,g \otimes g\ T\ (g\otimes g)^{-1} \mb{and} T \mapsto  P\ T^{t_1t_2}\,P\,,
\eeq
where $\lambda$ is a complex non-zero parameter, $g$ is any invertible $2\times2$ matrix and $P$ is the permutation, see \eqref{S-Scheck}.

\subsection{Case $\boldsymbol{m_{42}=0}$ and  $\boldsymbol{m_{24}\neq 0}$}

This case can be deduced from the case 
$m_{24}=0, m_{42}\neq 0$ in the following way. From any solution $\check S(z_1,z_2)$ to the braided Yang--Baxter equation, one can construct a new one 
\begin{equation}\label{eq:transf}
\check \Sigma(z_1,z_2)=\check S^{t_1t_2}(1/z_2,1/z_1)\;,
\end{equation}
where $(.)^{t_j}$ denotes the transposition in space $j$. Then, starting from 
\bea
\check S(z_1,z_2) &=& -\frac{z_2}{z_1}\,\Lambda(z_1,z_2)\,\Lambda(z_2,z_1)^{-1}\mb{with}
\Lambda(z_1,z_2) = T-\frac{m_{24}}{z_2}\,\II_4 
\eea
we get 
\bea
\check \Sigma(z_1,z_2) &=& - \frac{z_2}{z_1}\,\bar\Lambda(z_1,z_2)\,\bar\Lambda(z_2,z_1)^{-1}\mb{with}
 \bar\Lambda(z_1,z_2) = T^{t_1t_2}-m_{24}\, z_1\,\II_4 \; .\qquad
\eea
It is clear that, up to a replacement $m_{24} \leftrightarrow m_{42}$,  $\check \Sigma(z_1,z_2)$ corresponds to the expression of 
the S-matrix \eqref{eq:Smatrix} with $m_{24}=0$. 
Therefore, any solution $\check S(z_1,z_2)$ of the Yang--Baxter equation for $m_{42}=0$ is obtained 
from a solution $\check \Sigma(z_1,z_2)$ for $m_{24}=0$ (classified in the previous paragraph).

For the case $m_{42}=0$, the condition on $T$  reads
\begin{eqnarray}\label{eq:m42=0}
 T_{12} T_{23} T_{12}+ (T_{23})^2 \  T_{12} = T_{23} T_{12} T_{23}+ T_{23} \ (T_{12})^2
\end{eqnarray}
which is deduced from relation \eqref{eq:m24=0} by transposition, as expected.
One recognizes now in \eqref{eq:m42=0} the defining relations of the algebra $\cS_3$, see \cite{nous}.

\appendix

\section{Explicit expression of the solvable Hamiltonians}
Gathering the results presented above, we give here the explicit form of the Hamiltonians that we can solve.
We present them accordingly to the decomposition done in section \ref{sec:ybe}. 

To be compact, we first decompose  the Hamiltonian \eqref{eq:ham12} as
\beq
 h =\wt h +T_9 +m_{11}\,\II_9
\eeq
where the decomposition reflects the two steps used in the CBA. 
Then, using relations \eqref{1st.constraint} one gets
\beq
\begin{aligned}
&\wt h = \begin{pmatrix}
0 & 0 & 0 & 0 & 0 & 0 & 0 & 0 & 0 \\
0 & m'_{22} & m_{23} & m_{24} & 0 & 0 & 0 & 0 & 0 \\
0 & m_{32} & m'_{33} & 0 & 0 & 0 & m_{24} & 0 & 0 \\
0 & m_{42} & 0 & m'_{44} & 0 & 0 & -m_{23} & 0 & 0 \\
0 & 0 & 0 & 0 & -m'_{22}-m'_{44} & m_{23} & 0 & -m_{23} & 0 \\
0 & 0 & 0 & 0 & m_{32} & -m'_{33}-m'_{44} & 0 & 0 & -m_{23} \\
0 & 0 & m_{42} & -m_{32} & 0 & 0 & m'_{77} & 0 & 0 \\
0 & 0 & 0 & 0 & -m_{32} & 0 & 0 & -m'_{22}-m'_{77} & m_{23} \\
0 & 0 & 0 & 0 & 0 & -m_{32} & 0 & m_{32} & -m'_{22}-m'_{44}
\end{pmatrix}
\label{eq:ham-0}
\end{aligned}
\eeq
\beq
T_9 = 
\begin{pmatrix}
0 & 0 & 0 & 0 & 0 & 0 & 0 & 0 & 0 \\
0 & 0 & 0 & 0 & 0 & 0 & 0 & 0 & 0 \\
0 & 0 & 0 & 0 & 0 & 0 & 0 & 0 & 0 \\
0 & 0 & 0 & 0 & 0 & 0 & 0 & 0 & 0 \\
0 & 0 & 0 & 0 & t_{55} & t_{56} & 0 & t_{58} & t_{59} \\
0 & 0 & 0 & 0 & t_{65} & t_{66} & 0 & t_{68} & t_{69} \\
0 & 0 & 0 & 0 & 0 & 0 & 0 & 0 & 0 \\
0 & 0 & 0 & 0 & t_{85} & t_{86} & 0 & t_{88} & t_{89} \\
0 & 0 & 0 & 0 & t_{95} & t_{96} & 0 & t_{98} & t_{99}
\end{pmatrix}
\label{eq:T9}
\eeq
where  the $t_{ij}$ variables are the entries of the matrix $T$ given in \eqref{eq:Tmat} and  $m'_{jj}=m_{jj}-m_{11}$, $\forall j$. 

The $t_{ij}$ variables are constrained by solvability, and we give below their possible expressions, according to the classification done in this article. The remaining variables are free, but for the relation 
$$m'_{22}+m'_{44}=m'_{33}+m'_{77}.$$ 
This provides  7 free parameters, in addition to the free parameters included in $T$.

\subsection{Forms of $T$ for $\boldsymbol{m_{24}\,m_{42}\neq 0}$\label{sec:repHecke}}
The $4\times4$ representations of the Hecke algebra have being given\footnote{Be careful that the solutions in \cite{Hiet} have to be braided (i.e. multiplied by the permutation $P$ on the left) to get $\wt T$.} in section \ref{sec:Hecke}. 
They provide the form of $\wt T$ (up to the transformations \eqref{transfo:hecke}) which is related to $T$ via \eqref{eq:T-Ttilde}. 
Altogether, we get the following form for $T$ up to the transformations \eqref{transfo:hecke}.

\begin{equation}
\label{eq:soluHecke}
\begin{aligned}
&T_{H11}= \tau\,\begin{pmatrix} \sinh(\theta)+\eps &0 & 0& \sinh(\theta) \\
 0 & \sinh(\theta) & \cosh(\theta) & 0 \\ 0 & \cosh(\theta) & \sinh(\theta) & 0 \\ \sinh(\theta) & 0 & 0 & \sinh(\theta)-\eps
 \end{pmatrix}+\rho\,\II_4\mb{;}
 \\[1ex]
&T_{H12}= \tau\,\begin{pmatrix} a &0 & 0& b \\
 0 & 0 & a^{-1} & 0 \\ 0 & a & a-a^{-1} & 0 \\ 0 & 0 & 0 & -a^{-1} 
 \end{pmatrix}+\rho\,\II_4\mb{;}
T_{H13}= \eps\tau\,\begin{pmatrix} 1 &b & -b& ab \\
 0 & 0 & 1 & -a \\ 0 & 1 & 0 & a \\ 0 & 0 & 0 & 1 
 \end{pmatrix}+\rho\,\II_4\mb{;}
 \\[1ex]
&T_{H21/H22}= \tau\,\begin{pmatrix} a &0 & 0& 0 \\
 0 & 0 & b^{-1} & 0 \\ 0 & b & a-a^{-1} & 0 \\ 0 & 0 & 0 & \eps\,a^\eps 
 \end{pmatrix}+\rho\,\II_4\mb{;}
 T_{H14}= \tau\,\begin{pmatrix} 0 &0 & 0& a \\
 0 & 0 & \eps & 0 \\ 0 & \eps & 0 & 0 \\ a^{-1} & 0 & 0 & 0
 \end{pmatrix}+\rho\,\II_4
\\[1ex]
&
T_{H03}= (\tau+\rho)\,\II_4,
\end{aligned}
\end{equation}
where $a$, $b$ and $\theta$ are free complex parameters and $\eps=\pm1$, while $\tau$ and $\rho$ are given in \eqref{eq:mutau} with the change of variables:
\begin{eqnarray}
R_{H11} &:& \exp\theta = 2p^2\mb{and} q=\frac{\eps}{2p}
\\
R_{H12} &:& a=p\,,\quad q=\frac{1}{a}\mb{and} k=b
\\
R_{H13} &:& \eps=k^2\,,\quad a=\eps kq\mb{and} b=\eps kp
\\
R_{H21/H22} &:& a=k^2\mb{and} kp=b
\\
R_{H14} &:& a=p.
\end{eqnarray}

We remind that $P$ is the permutation, see \eqref{S-Scheck}. 
The index on $T$ refers to the solution in \cite{Hiet} it corresponds to, see table \ref{hieta}. 

\subsection{Forms of $T$ for $\boldsymbol{m_{24}\neq 0}$ and  $\boldsymbol{m_{42}= 0}$\label{sec:repSn}}
The $4\times4$ representations of the $\cS_n$ algebra have the following form up to the transformations given in \eqref{eq:transfoTn}

\begin{equation}
\label{eq:soluSn}
\begin{aligned}
&T_{(1)}=\begin{pmatrix}
                0 & 0 & 0 & 0 \\
                b & c & 0 & 0 \\
                d & 0 & 0 & 0 \\
                0 & a & 0 & 0 
               \end{pmatrix} \ {;}\
T_{(2)}=\begin{pmatrix}
                0 & b & c & d \\
                0 & 0 & 0 & a \\
                0 & 0 & 0 & {b+c-a} \\
                0 & 0 & 0 & 0
               \end{pmatrix}
\ {;}\
T_{(3)}=\begin{pmatrix}
          0 & 0 & 0 & 0 \\
          b & \frac{ab}{c} & 0 & 0 \\
          c & 0 & a & 0 \\
          0 & 0 & 0 & 0
         \end{pmatrix},
\quad\\[1ex]
&T_{(4)}=\begin{pmatrix}
          0 & 0 & 0 & 0 \\
          0 & b & 0 & 0 \\
          0 & c & 0 & 0 \\
          0 & 0 & 0 & a
         \end{pmatrix}
\mb{;}
T_{(5)}=\begin{pmatrix}
          a & 0 & 0 & 0 \\
          0 & b & 0 & 0 \\
          0 & 0 & c & 0 \\
          0 & 0 & 0 & 0
         \end{pmatrix} \mb{;} 
T_{(6)}=\begin{pmatrix}
          0 & 0 & 0 & 0 \\
          b & c & 0 & a \\
          0 & 0 & 0 & 0 \\
          0 & -b & b & 0 
         \end{pmatrix}, \quad 
\\[1ex]
&T_{(7)}=\begin{pmatrix}
          -a & 0 & 0 & 0 \\
          b & 0 & a & 0 \\
          0 & 0 & -a & 0 \\
          0 & 0 & 0 & 0 
         \end{pmatrix},
\end{aligned}
\end{equation}
where $a$, $b$, $c$ and $d$ are complexe free parameters. 
The index on $T$ refers to the solutions of the classification done in \cite{nous}. 
\subsection{Forms of $T$ for $\boldsymbol{m_{24}=0}$ and  $\boldsymbol{m_{42}\neq 0}$.}
The $4\times4$ representations of the $\cT_n$ algebra are obtained from the ones of the $\cS_n$ algebra by a simple transposition, 
see section \ref{sec:repSn}.  We don't repeat them here.

\subsection{Reshetikhin criterion}
The Reshetikhin criterion \cite{KS-resh} is defined as
\beq\label{eq:Resh_crit}
{\big[h_{j,j+1}+h_{j+1,j+2}\,,\,[h_{j,j+1}\,,\,h_{j+1,j+2}]\big]} \,=\, A_{i+1,i+2}-A_{i,i+1}
\eeq
where $h$ is the Hamiltonian, and $A$ some matrix in $End(\CC^3\otimes\CC^3)$ to be determined. When obeyed, it ensures that for a given Hamiltonian $h$, an $R$-matrix exists and satisfies the Yang-Baxter relation in its difference form.
It is a necessary condition when one assumes that the Hamiltonian $h$ is deduced from an $R$-matrix, that is moreover supposed to depend only on the difference (or the ratio) of the two spectral parameters. 

Let us stress that the Reshetikhin criterion may fail if $h$ is obtained from a Lax matrix that is not directly constructed from $R(u)$, as for instance in \cite{Mart}, or if $R$ depends separately on the two spectral parameters, as it is the case for the Hubbard model.

This criterion is invariant under shifts by the identity matrix, conjugation and transposition at the level of the $9\times9$ matrix $h$. 
Nevertheless, one should restore the transformations \eqref{transfo:hecke} or \eqref{eq:transfoTn} at the level of the matrix $T$, before testing it. However, one can check that the $4\times4$ conjugaison on $T$ is equivalent to a $9\times 9$ conjugaison on $T_9$.

We  looked for an $A$-matrix (see \eqref{eq:Resh_crit}) and a dilatation $\lambda$ 
(see \eqref{transfo:hecke} or \eqref{eq:transfoTn}) such that the criteria is fulfilled for the Hamilonians $ h$, 
built on the matrices $T$ given in \eqref{eq:soluHecke} (and also for $P.T.P$ or $T^{t_1t_2}$) or on the matrices $T$ 
given in \eqref{eq:soluSn} (and also for $P.T^{t_1t_2}.P$).
 None of the Hamiltonians presented here obey this criterion (unless one constrains some of the free parameters of $ h$).
 
This indicates that if it exists, the $R$-matrix either depends on the two spectral parameters, or that the Hamiltonian is obtained from a Lax matrix that is not an $R$-matrix. Remark that the $4\times4$ $R$-matrix associated to the  $\cS_n$ or $\cT_n$ algebras indeed depends separately on the two spectral parameters.

\end{document}